\documentclass[twocolumn,showpacs,showkeys,preprintnumbers,amsmath,amssymb]{revtex4-1}
\usepackage{bbm}
\usepackage{amsfonts}

\usepackage{epsfig}
\usepackage{graphicx}
\usepackage{amssymb}   
\usepackage{dcolumn}
\usepackage{bm}
\usepackage[colorlinks,citecolor=blue, linkcolor=blue,hyperindex,CJKbookmarks]{hyperref}
\begin{document}
\vspace{2.5cm}
\title{Robust quantum state transfer between two superconducting qubits via partial measurement}
\author{Yan-Ling Li$^{1}$}

\author{Yao Yao$^{2}$}
\altaffiliation{yaoyao@mtrc.ac.cn}
\author{Xing Xiao$^{3}$}
\altaffiliation{xiaoxing1121@gmail.com}
\affiliation{$^{1}$School of Information Engineering, Jiangxi University of Science and Technology, Ganzhou, Jiangxi 341000, China\\
$^{2}$Microsystems and Terahertz Research Center, China Academy of Engineering Physics, Chengdu, Sichuan 610200, China\\
$^{3}$College of Physics and Electronic Information, Gannan Normal University, Ganzhou Jiangxi 341000, China
}

\begin{abstract}
We develop a potentially practical proposal for robust quantum state transfer (QST) between two superconducting qubits coupled by a coplanar waveguide (CPW) resonator. We show that the partial measurement could drastically enhance the fidelity even when the dissipation of qubits and CPW is considered. Unlike many other schemes for QST, our proposal does not require the couplings between the qubits and the CPW resonator to be strong. In fact, our method works much more efficiently in the weak coupling regime. The underlying mechanism is attributed to the probabilistic nature of partial measurement.

\end{abstract}
\maketitle

\section{Introduction}
\label{intro}
The ability to coherently transfer quantum information between a number of qubits plays a crucial role in quantum computation and quantum communication \cite{bennett2000}. The importance of this procedure stems from the fact that the QST between distant qubits, together with some other local quantum processings, not only support the possibility for a variety of novel applications, such as quantum key distribution \cite{shor2000} and quantum teleportation \cite{bennett1993}, but also would be beneficial for large-scale quantum devices and distributed quantum computation. 

In the past two decades, QST has attracted considerable attentions and been studied both theoretically and experimentally in various physical systems ranging from linear optical systems \cite{matsuk2004,stute2013}, trapped atoms and ions \cite{cirac1997,yin2007}, optomechanical systems \cite{wang2012a,wang2012b}, spin chain \cite{shi2005,yao2011}, superconducting circuits \cite{lyakhov2005,sillanpaa2007}, nuclear magnetic resonance \cite{zhang2006}, nitrogen-vacancy centers \cite{yang2011} and some other hybrid systems \cite{singh2012}. Many very interesting methods have been proposed to realize the perfect QST,  such methods include: pre-engineering the interqubit couplings \cite{christandl2004},  dynamically varying the coupling constants between the first and the last pair of qubits \cite{lyakhov2006}, virtually exciting the transmission channel \cite{plenio2005,liyong2005}, dual-rail encoding by two parallel quantum channels \cite{burgarth2005a,burgarth2005b}, utilizing adiabatic passage technique \cite{eckert2007}, and by weak measurement \cite{he2013,man2014}.
However, the perfect QST is still faced with many challenges in the experiment, since the fidelity of the state transfer is highly limited by the decoherence of qubits, the quality of the transmission channel, and the imperfections during the experimental operations. This would be the most limiting factor for the applications of QST in quantum information processing.
In this context, it is an extremely important issue to ensure the faithful transfer of quantum state in the realistic scenarios.

In this paper, we propose a scheme to improve the fidelity of QST with the assistance of quantum partial measurement. Partial measurement which is also known as weak measurement (note the difference with weak value), is a generalization of von
Neumann measurement. By ``partial'' or ``weak'', we mean that the quantum state doesn't compleltely collapse towards an eigenstate of the measurement operator, and could be reversed with some operations. In some sense, partial measurement is more valuable than von Neumann measurement, because numerous researches have indeed demonstrated that partial measurement could be used to steer the quantum state evolution \cite{katz06,blok2014}, protect quantum entanglement and quantum correlations \cite{sun09,man12,xiao13,katz08,kim09,kim12}, and enhance the teleportation of quantum Fisher information \cite{xiao2016} in amplitude damping decoherence. This motivates us to study the QST under decoherence by utilizing the partial measurement. 

The key point of our paper is that partial measurement is nondestructive and can be reversed with a certain probability, such that one can probabilistically project the state towards ground state which is insensitive to decoherence by a pre-partial measurement, and then retrieve the initial state with a partial measurement reversal. This procedure is similar to the way of ``conclusive transfer'' proposed by Burgarth and Bose \cite{burgarth2005a}, which means that the receiver has a certain probability for obtaining the perfectly transferred state. However, we should point out that our scheme is essentially different with that proposed in \cite{burgarth2005a}, where the method of dual-rail encoding is involved. Our scheme is more simple and practical. Remarkably, our method is robust to the strength of qubit decoherence, the coupling strength and the quality of transmission channel.  

\begin{figure}
  \includegraphics[width=0.5\textwidth]{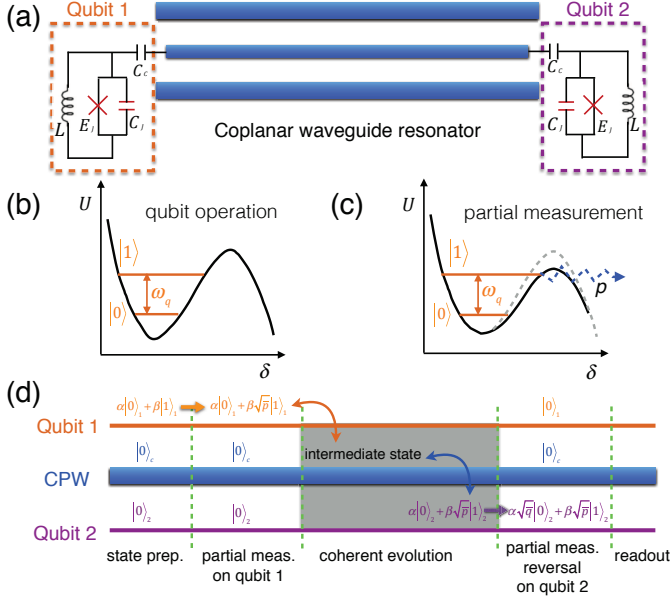}
\caption{(color online) (a) The schematic of quantum state transfer setup between two phase qubits coupled via a coplanar waveguide resonator. (b) Potential-energy diagram of the phase qubit, where $\delta$ is the superconducting phase difference across the Josephson junction. The phase qubit is formed by the two lowest eigenstates $|0\rangle$ and $|1\rangle$ with the transition frequency $\omega_{q}$. (c) Illustration of the partial measurement on the phase qubit. By a partial measurement pulse, the state $|1\rangle$ tunnels out of the well with probability $p$ which is tunable by changing the amplitude of the measurement pulse. Note that the probability of tunneling out state $|0\rangle$ is much smaller than (typically $\sim1/1000$) that of state $|1\rangle$ \cite{martinis2002}, thus it can be safely neglected. (d) Illustration of the ideal process of QST by partial measurement and partial measurement reversal. }
\label{Fig1}       
\end{figure}

\section{Physical Model}

Circuit quantum electrodynamics (circuit QED) system is a particularly suitable platform to demonstrate our proposal, not only due to the paramount advantage of such system for the ability to fabricate, manipulate and process quantum information tasks in a scalable way, but also due to the excellent feasibility of the partial measurement in a superconducting phase qubit. Notice that the present scheme is also valid for other types of qubit but may require more complicated operations. In fact, the partial measurement is a particular type of POVM formalism which could be realized with the combination of projective measurement and unitary operation of a composite system including ancillary system and target system. Hence, the performance of partial measurement on the target qubit is equivalent to the action of von Neumann projective measurement on the ancilla qubit which is previously coupled to it.

We consider the setup schematically illustrated in figure \ref{Fig1}(a), where two phase qubits with transition frequency $\omega_{q}$ interact resonantly with a CPW resonator, i.e., $\omega_{q}=\omega_{c}$. Under the rotating wave approximation and in the interaction picture, the Hamiltonian of the system can be written as ($\hbar=1$),
\begin{equation}
H=g_{1}(a\sigma_{1}^{+}+a^{\dagger}\sigma_{1}^{-})+g_{2}(a\sigma_{2}^{+}+a^{\dagger}\sigma_{2}^{-}),
\label{eq1}
\end{equation}
where $\sigma_{+}$ and $\sigma_{-}$ are the up and down operators, $a^{\dagger}$ and $a$ are the creation and annihilation operators of the CPW resonator, and $g_{n}$ ($n=1$, $2$) is the coupling constant between the \emph{n}th qubit and the CPW resonator.

Based on this model, we want to transfer the quantum information coded in the first phase qubit to the second one. This means to accomplish the following process,
$\left(\alpha|0\rangle_{1}+\beta|1\rangle_{1}\right)|0\rangle_{2} \rightarrow |0\rangle_{1}\left(\alpha|0\rangle_{2}+\beta|1\rangle_{2}\right)$, where $\alpha^2+|\beta|^2=1$ ($\alpha$ is a real number). However, any quantum system inevitably interacts with the environment which leads to the decoherence, which reduces the fidelity of QST. Here, we consider the scheme of using partial measurement and its reversal to suppress the decoherence and improve the fidelity.
\section{Robust QST by partial measurements}

Before the demonstration of our scheme, we should review the concepts of partial measurement and partial measurement reversal.
Partial measurement, unlike the standard von Neumann projective measurement projecting the initial state to one of the eigenstates of the measurement operator, slightly disturbs the state, only extracts partial information and consequently yields a nonunitary, nonprojective transformation of the quantum state. For a single qubit with computational basis $|0\rangle$ and $|1\rangle$, the so-called partial measurement is
\begin{eqnarray}
\label{e2}
\mathcal{M}_{0}&=&|0\rangle\langle0|+\sqrt{1-p}|1\rangle\langle1|,\\
\label{e3}
\mathcal{M}_{1}&=&\sqrt{p}|1\rangle\langle1|,
\end{eqnarray}
where $\mathcal{M}_{0}^{\dagger}\mathcal{M}_{0}+\mathcal{M}_{1}^{\dagger}\mathcal{M}_{1}=I$ and the parameter $p$ $(0\leqslant p\leqslant1)$ is usually named as the strength of partial measurement. Note that $\mathcal{M}_{0}$ and $\mathcal{M}_{1}$ are not necessarily projectors and also nonorthogonal to each other. As shown in figure~\ref{Fig1}(c), the process of the state $|1\rangle$ tunneled out corresponds to the measurement operator $\mathcal{M}_{1}$, which is the same as von Neumann projective measurement and associated with an irrevocable collapse. However,  if the state $|1\rangle$ is not tunneled out, then $\mathcal{M}_{0}$ is a partial measurement which could be reversed if $p\neq1$.

The reverse procedure can be described by a non-unitary operator
\begin{eqnarray}
\label{e4}
\mathcal{M}_{0}^{-1}&=&\frac{1}{\sqrt{1-q}}\left(\begin{array}{cc}0 & 1 \\1 & 0\end{array}\right)\left(\begin{array}{cc}1 & 0 \\0 & \sqrt{1-q}\end{array}\right)\left(\begin{array}{cc}0 & 1 \\1 & 0\end{array}\right),\\
&=&\frac{1}{\sqrt{1-q}}X\mathcal{M}_{0}X,\nonumber
\end{eqnarray}
where $X=|0\rangle\langle1|+|1\rangle\langle0|$ is the qubit bit-flip operation (i.e., Pauli X operation). Although the non-unitary operator of equation (\ref{e4}) cannot be realized as an evolution with a suitable Hamiltonian, it can be achieved with certain probability by using another partial measurement. The second line of equation (\ref{e4}) indicates that the reversal process could be constructed by three steps: a bit-flip operation (i.e., a $\pi$-pulse in circuit QED experiments), a second partial measurement and a second bit-flip operation.
The factor $1/\sqrt{1-q}$ is related to the non-unitary nature of the partial measurement, i.e., probabilistic.
Therefore, the partial measurement reversal $\mathcal{M}_{0}^{-1}$ could exactly undo the partial measurement $\mathcal{M}_{0}$ in a probabilistic way by properly choosing the parameter $q$.

Following the circuit illustrated in figure \ref{Fig1}(d), we show that the partial measurement could effectively enhance the fidelity without regard to the decoherence and coupling strength.
Initially, assuming the total system is in the state
\begin{equation}
|\psi(0)\rangle=\left(\alpha|0\rangle_{1}+\beta|1\rangle_{1}\right)|0\rangle_{2}|0\rangle_{c},
\label{eq5}
\end{equation}
where $|0\rangle_{c}$ indicates the CPW resonator in the vacuum state. Before transferring the quantum state, a partial measurement is performed to the first qubit. Then the state of the system is changed to (with respect to the no tunneling of state $|1\rangle$)
\begin{equation}
|\psi(0)\rangle=\frac{1}{N_{1}}\left(\alpha|0\rangle_{1}+\beta\sqrt{1-p}|1\rangle_{1}\right)|0\rangle_{2}|0\rangle_{c},
\label{eq6}
\end{equation}
where the normalization factor $N_{1}=\sqrt{\alpha^{2}+|\beta|^{2}(1-p)}$, $p$ is the strength of partial measurement. Remarkably, even though the state is not actually measured (no tunneling happened), the amplitude of
the state $|0\rangle$ is amplified due to the shrink of the normalization factor. As $p$ approaches unity, the state $|\psi(0)\rangle$ is almost completely projected into the state $|0\rangle$, which is immune to dissipative decoherence.

Considering the spontaneous emission of the qubits and the decay of the CPW resonator, the time evolution of the state is dominated by the master equation.
\begin{eqnarray}
\label{eq7}
\frac{\partial}{\partial t}\rho &=& -i[H,\rho]-\frac{\kappa}{2}(a^{\dagger}a\rho+\rho a^{\dagger}a-2a\rho a^{\dagger})\\
&&-\sum_{n=1,2}\frac{\Gamma_{n}}{2}(\sigma_{n}^{+}\sigma_{n}^{-}\rho+\rho\sigma_{n}^{+}\sigma_{n}^{-}-2\sigma_{n}^{-}\rho\sigma_{n}^{+}), \nonumber
\end{eqnarray}
where $\kappa$ is the decay constant of the CPW resonator, $\Gamma_{1}$ and $\Gamma_{2}$ are the spontaneous emission rates of the first and the second qubit, respectively.
Since no more than one excitation is involved in the whole system, the subspace is spanned by $\{|1\rangle = |0\rangle_{1}|0\rangle_{2}|0\rangle_{c}$, $|2\rangle = |1\rangle_{1}|0\rangle_{2}|0\rangle_{c}$, $|3\rangle = |0\rangle_{1}|0\rangle_{2}|1\rangle_{c}$, $|4\rangle = |0\rangle_{1}|1\rangle_{2}|0\rangle_{c}\}$. Then the equation (\ref{eq7}) could be analytically solved. The time evolution of the density matrix is given as
\begin{eqnarray}
\rho(t)= \left(\begin{array}{cccc}
\rho_{11} & \rho_{12} & \rho_{13}& \rho_{14}\\
 \rho_{21} & \rho_{22} & \rho_{23}& \rho_{24}\\
  \rho_{31} & \rho_{32} & \rho_{33}& \rho_{34}\\
   \rho_{41} & \rho_{42} & \rho_{43}& \rho_{44}\end{array}\right),
\label{eq8}
\end{eqnarray}
where
\begin{eqnarray}
\rho_{11}&=& 1+(a-1)e^{-s t},\nonumber\\
\rho_{12}&=& \rho_{21}^{*}= \frac{c}{r^{2}}[g_{1}^{2}\cos(rt)+g_{2}^2]e^{-\frac{s t}{2}},\nonumber\\
\rho_{13}&=& \rho_{31}^{*}= ic\frac{g_{1}}{r}\sin(rt)e^{-\frac{s t}{2}},\nonumber\\
\rho_{14}&=& \rho_{41}^{*}= c\frac{g_{1}g_{2}}{r^{2}}[\cos(rt)-1]e^{-\frac{s t}{2}},\nonumber\\
\rho_{22}&=& \frac{b}{2r^{4}}[g_{1}^{4}+2g_{2}^{4}+g_{1}^{4}\cos(2rt)+4g_{1}^{2}g_{2}^{2}\cos(rt)]e^{-s t},\nonumber\\
\rho_{23}&=& \rho_{32}^{*}=ib\frac{g_{1}}{2r^{3}}[g_{1}^{2}\sin(2rt)+2g_{2}^{2}\sin(rt)]e^{-s t},\nonumber\\
\rho_{24}&=& \rho_{42}^{*}=b\frac{g_{1}g_{2}}{2r^{4}}[g_{1}^{2}-2g_{2}^{2}\nonumber\\
&&+2(g_{2}^{2}-g_{1}^{2})\cos(rt)+g_{1}^{2}\cos(2rt)]e^{-st},\nonumber\\
\rho_{33}&=& b\frac{g_{1}^{2}}{2r^{2}}[1-\cos(2rt)]e^{-st},\\
\rho_{34}&=& \rho_{43}^{*} = ib\frac{g_{1}^{2}g_{2}}{2r^{3}}[2\sin(rt)-\sin(2rt)]e^{-st},\nonumber\\
\rho_{44}&=& b\frac{g_{1}^{2}g_{2}^{2}}{2r^{4}}[\cos(2rt)-4\cos(rt)+3]e^{-st},\nonumber
\label{eq9}
\end{eqnarray}
with $r=\sqrt{g_{1}^{2}+g_{2}^{2}}$, $a= \alpha^{2}/N_{1}^{2}$, $b= |\beta|^{2}(1-p)/N_{1}^{2}$ and $c=\alpha \beta^{*} \sqrt{1-p}/N_{1}^{2}$. In order to to concentrate on the basic idea of our scheme and simplify the analytic expressions, we have assumed $\kappa=\Gamma_{1}=\Gamma_{2}=s$. However, our proposal is also feasible for the most general situations $\kappa\neq\Gamma_{1}\neq\Gamma_{2}$.

Finally, a partial measurement reversal and a $\sigma_{z}$ operation are added to the second qubit.
\begin{eqnarray}
\rho(t)= \frac{1}{N} \left(\begin{array}{cccc}
\rho_{11}\overline{q} & \rho_{12}\overline{q} & \rho_{13}\overline{q}& -\rho_{14} \sqrt{\overline{q}}\\
 \rho_{21}\overline{q} & \rho_{22}\overline{q} & \rho_{23}\overline{q}& -\rho_{24}\sqrt{\overline{q}}\\
  \rho_{31}\overline{q} & \rho_{32}\overline{q} & \rho_{33}\overline{q}& -\rho_{34}\sqrt{\overline{q}}\\
   -\rho_{41}\sqrt{\overline{q}} & -\rho_{42} \sqrt{\overline{q}}& -\rho_{43}\sqrt{\overline{q}}& \rho_{44}\end{array}\right)
   \label{eq10}
\end{eqnarray}
where $\overline{q}=1-q$, $q$ is the strength of the partial measurement reversal and the normalization factor $N=1-q+q\rho_{44}$.

\begin{figure}
  \includegraphics[width=0.5\textwidth]{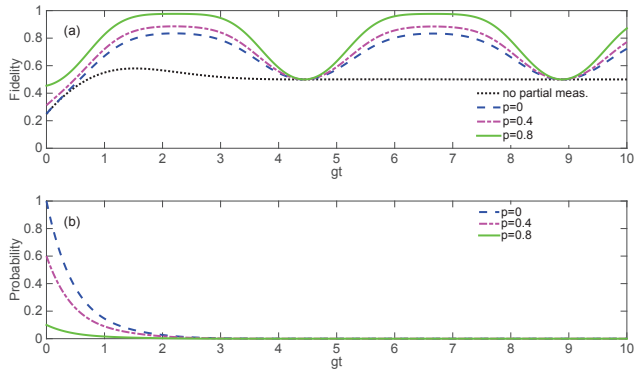}
\caption{(color online) The plots of the fidelity (a)  and the success probability (b) as a function of scaled time $gt$ under various partial measurement strength $p$ in the weak coupling regime ($g=0.5s$ with $g_{1} = g_{2} = g$). The other parameters are $\alpha=\beta=1/\sqrt{2}$, and the partial measurement reversal $q=1-(1-p)\exp(-st)$. }
\label{Fig2}       
\end{figure}

At time $t$ and the fidelity of state $\alpha|0\rangle+\beta|1\rangle$ transferring from the first qubit to the second qubit is defined as
\begin{eqnarray}
F&=&\langle\varphi|\rho(t)|\varphi\rangle\\
&=&\frac{1}{N}\left[\alpha^{2}\overline{q}\rho_{11}-\alpha \beta \rho_{14} \sqrt{\overline{q}}-\alpha \beta^{*} \rho_{41} \sqrt{\overline{q}}+ |\beta|^{2}\rho_{44}\right],\nonumber
\label{eq11}
\end{eqnarray}
where $|\varphi\rangle = |0\rangle_{1}(\alpha|0\rangle_{2}+\beta|1\rangle_{2})|0\rangle_{c}$ is the target state. Following the quantum jump method demonstrated in Ref. \cite{korotkov2010}, the optimal reversing measurement strength is calculated to be (see details in the Appendix)
\begin{equation}
q=1-(1-p)e^{-st}.
\label{eq12}
\end{equation} 
Even though equation (\ref{eq12}) is not the most optimal one, it is concise and applicable since it is state-independent (i.e., without involving the initial parameters $\alpha$ and $\beta$). Note that one can obtain the most optimal reversing measurement strength that gives the maximal fidelity by utilizing partial differentiation to optimize the variable parameter $q$, but the generally analytic expression is too complicated to present, particularly, it is state-dependent.

To demonstrate the power of partial measurement we plot the fidelity of QST as a function of dimensionless time $gt$ in figure~\ref{Fig2}(a), where we have assumed $g_{1} = g_{2} = g$ and in the weak coupling regime $g=0.5s$. We see that the fidelity decays very fast under the influence of qubit decoherence and cavity loss (dotted line). However, we note that only partial measurement reversal (i.e., $p=0$) could effectively enhance the fidelity. Remarkably, the fidelity can be further improved with the combination of partial measurement and its reversal. It is possible to eliminate the decoherence effects completely provided that the strength of the partial measurement $p$ and that of the corresponding optimal reversing measurement $q$
are sufficiently strong. Since a partial measurement
is equivalent to a probabilistic state rotation with the rotation angle corresponding to the strength of the measurement, the price of high fidelity of QST is based on the low success probability. The trade-off between the high fidelity and low success probability needs careful optimization. For our approach, the success probability is $P=1-q+q\rho_{44}$.
As shown in figure~\ref{Fig2}(b), the higher the fidelity, the larger the partial measurement strength $p$, and hence the lower success probability $P$.
\begin{figure}
  \includegraphics[width=0.5\textwidth]{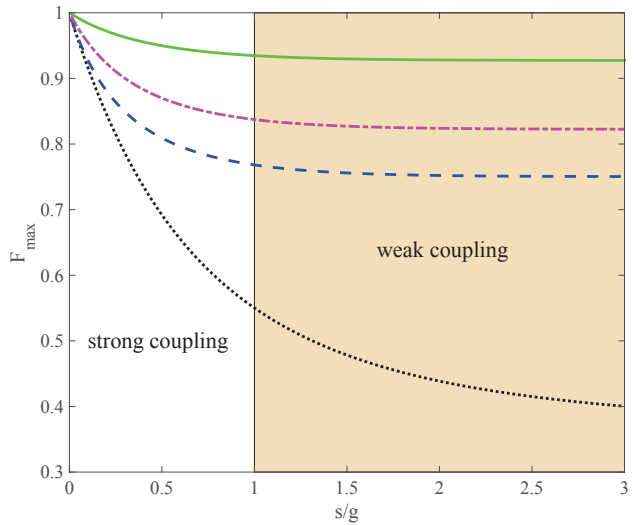}
\caption{(color online) The maximal achievable fidelities $F_{\max}$ as a function of scaled decay rate $s/g$. The shadow regions $s/g>1$ indicate the weak coupling regime. The dotted line shows the result of natural evolution, and the dashed line ($p=0$), dotted-dashed line ($p=0.4$) and solid line ($p=0.8$). The other parameters are $\alpha=0.6$, $\beta=0.8$, and $q=1-(1-p)\exp(-st)$.}
\label{Fig3}       
\end{figure}

Even more remarkably, we find that the partial measurement assisted QST is not only robust against the decoherence effects, but also resilient to the coupling strength $g$. 
In figure~\ref{Fig3}, we show the maximally achievable fidelity for different scenarios. As we all know, the smaller the coupling strength $g$ is, the longer the time is required for coherent transfer between the qubit and CPW. Thus the energy dissipation plays a dominant role in the procedure of QST, which inevitably reduces the maximally achievable fidelity. It is immediate to note that the maximal fidelity degrades fast in the weak coupling regime if no partial measurement is applied. 
However, the partial measurement can greatly prevent the degradation and enable the maximal fidelity to be robust with respect to the increase of decay rate. The crucial fact is that partial measurement intentionally moves the qubit close to its ground state. Energy dissipation is naturally suppressed in this inactive state, and the quantum state could be preserved for a long time with little deterioration (without considering the pure dephasing decoherence, which will be discussed in the next section). Therefore, the influences of large energy dissipation and small coupling strength is suppressed. 
This result is not trivial since it implies that the proposal does not depend on the strong coupling between the qubit and the CPW resonator, as such it could greatly releases the experimental requirements.
However, in the small-$g$ limit (i.e., $s/g\gg1$), the operational time is so long that the quantum state is completely decayed before the coherent transfer. Then the fidelity approaches to $\alpha^2$. This value corresponds to the case of complete decoherence.

By comparing these results, it is easy to conclude
that the strength of partial measurement plays a central role for the robust QST. With the increase of $p$, the fidelity increases and becomes more robust against the decay rates and the coupling strength. The underlying physics could be understood as follows: from equation (\ref{eq6}), we know that the stronger the partial measurement strength $p$, the weight of the $|0\rangle$ state $\alpha/N_{1}$ increases, i.e., the closer the initial state is steered towards the $|0\rangle$ state. Then the state of equation (\ref{eq6}) becomes more insensitive to the qubit dissipation and the loss of CPW resonator. 
After the coherent evolution, an optimal partial measurement reversal is performed on the second qubit to retrive the initial state. Therefore, the fidelity of QST is not sensitive to the decoherence factors $\kappa$ and $\Gamma_{n}$, but rather highly depends on the partial measurement strength $p$. The perfect QST could be realized by the combination partial measurement and partial measurement reversal when $p\rightarrow1$.

\begin{figure}
  \includegraphics[width=0.5\textwidth]{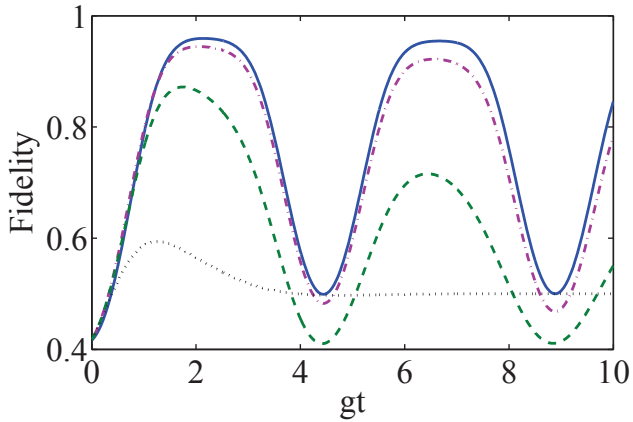}
\caption{(color online) Fidelity as a function of the scaled time $gt$ when both energy dissipation and qubit dephasing are considered. The lines from top to bottom correspond to different qubit dephasing rates: solid line ($\Gamma_{\phi}=0$), dotted-dashed line ($\Gamma_{\phi}=0.01g$), dashed line ($\Gamma_{\phi}=0.1g$) and dotted line ($\Gamma_{\phi}=g$). The partial measurement strength is chosen as $p=0.8$ and the other parameters are the same as figure \ref{Fig2}(a).}
\label{Fig4}       
\end{figure}
Before conclusion, we would like to point out that the present scheme is not valid for suppressing the pure dephasing decoherence. We confirm this result by numerical simulation. When the qubit dephasing is involved, the master equation of equation (\ref{eq7}) will include two pure dephasing terms of the qubits $\sum_{n=1,2}\frac{\Gamma_{\phi}}{2}(\sigma_{n}^{z}\rho\sigma_{n}^{z}-\rho)$, where $\Gamma_{\phi}$ is the dephasing rate of two phase qubits.
This master equation can be solved numerically using the Quantum Optics
Toolbox \cite{tan1999}. As shown in figure~\ref{Fig4}, we find that the fidelity of QST decreases with the increasing dephasing rate. This result is understandable because the dephasing is another decoherence mechanism which can be thought of as noise in the relative phase between the two qubit states $|0\rangle$  and $|1\rangle$. Projecting the initial states into $|0\rangle$ cannot remove the influence of dephasing. Hence, we expect that the existence of dephasing noise will reduce the maximally achievable fidelity by partial measurement.

\section{Conclusions}
\label{sec:4}

In conclusion we have proposed a scheme to realize the QST between two superconducting phase qubits. We demonstrated its advantages by performing a partial measurement on the first qubit while another partial measurement reversal is acted on the second qubit. The probabilistic nature of partial measurement enables the fidelity of QST to be greatly improved even considering the decay of qubits and CPW resonator, since the fidelity is no longer reliant on the decoherence factors $\kappa$ and $\Gamma_{n}$, but rather highly depends on the partial measurement strength $p$. We also show that our scheme is particularly robust to the coupling strength between the qubits and the CPW resonator. Namely, it works well in both strong and weak coupling regimes. 
Since any scheme for QST will suffer from some imperfections when implemented, the partial measurement is a powerful technique to overcome the decrease of fidelity.
%

\begin{acknowledgements}
This work is supported by the Funds of the National Natural Science
Foundation of China under Grant Nos. 11665004 and 11365011, and supported by Scientic Research Foundation of Jiangxi Provincial Education Department under Grants Nos. GJJ150996 and GJJ150682. Y. L. Li is supported by the Program of Qingjiang Excellent Young Talents, Jiangxi University of Science and Technology.
\end{acknowledgements}

\appendix
\section{optimal reversing measurement strength}

The optimal strength of measurement reversal of equation (\ref{eq12}) could be obtained from the perspective of quantum jump. Generally speaking, the initially pure state (\ref{eq5}) inevitably evolves into a mixed state because of the spontaneous emission of the qubits and the decay of the CPW resonator.
However, one can technically use the mathematical trick of ``unraveling'' the dissipation into ``jump'' and ``no jump'' scenarios and work with pure states \cite{scully1997}. Here, ``jump'' means the excited state $|1\rangle$ decays into the ground state $|0\rangle$, while ``no jump'' indicates the excited state $|1\rangle$ doesn't decay.

The brief sketch could be divided into three steps: 

(i) A partial measurement with strength $p$ is performed on the first qubit. 

(ii) After the evolution time $\sqrt{2}gt=\pi$, if the ``no jump'' cases occur, then the system stays in the superposition state of $|0\rangle_{1}|0\rangle_{2}|0\rangle_{c}$ and $|0\rangle_{1}|1\rangle_{2}|0\rangle_{c}$. Note that the amplitudes of the states $|1\rangle_{1}|0\rangle_{2}|0\rangle_{c}$ and $|0\rangle_{1}|0\rangle_{2}|1\rangle_{c}$ vanish at this time (see the $\rho_{22}$ and $\rho_{33}$ in the main text). 

(iii) The procedure of partial measurement reversal with strength $q$ and $\sigma_{2}^{z}$ is acted on the second qubit. 

The mathematical formulation is as follows (without normalization):
\begin{eqnarray}
&&(\alpha|0\rangle_{1}+\beta|1\rangle_{1})|0\rangle_{2}|0\rangle_{c},\nonumber\\
&&\xrightarrow{\text{(i)}}(\alpha|0\rangle_{1}+\beta\sqrt{\overline{p}}|1\rangle_{1})|0\rangle_{2}|0\rangle_{c},\\
\label{eqA1}
&&\xrightarrow{\text{(ii)}}\alpha|0\rangle_{1}|0\rangle_{2}|0\rangle_{c}-\beta\sqrt{\overline{p}}e^{-\frac{st}{2}}\alpha|0\rangle_{1}|1\rangle_{2}|0\rangle_{c},\\
\label{eqA2}
&&\xrightarrow{\text{(iii)}}\alpha\sqrt{\overline{q}}|0\rangle_{1}|0\rangle_{2}|0\rangle_{c}+\beta\sqrt{\overline{p}}e^{-\frac{st}{2}}\alpha|0\rangle_{1}|1\rangle_{2}|0\rangle_{c}.
\label{eqA3}
\end{eqnarray}

In order to recover the target state $|0\rangle_{1}(\alpha|0\rangle_{2}+\beta|1\rangle_{2})|0\rangle_{c}$ from the equation (\ref{eqA3}), it is easy to note that in the ``no jump'' case, the strength of the second partial measurement should satisfy the condition
\begin{equation}
\overline{q}=\overline{p}e^{-st}.
\end{equation}
Then the equation (\ref{eqA3}) reduces to the target state $|0\rangle_{1}(\alpha|0\rangle_{2}+\beta|1\rangle_{2})|0\rangle_{c}$ except for a constant factor which is determined by the probabilistic nature of the 
scheme.



\end{document}